\documentclass[%
 reprint,
 amsmath,amssymb,
 aps,
]{revtex4-2}

\usepackage{graphicx}% Include figure files
\usepackage{dcolumn}% Align table columns on decimal point
\usepackage{bm}% bold math
\usepackage{hyperref}% add hypertext capabilities
\usepackage[mathlines]{lineno}% Enable numbering of text and display math

\usepackage{multirow}
\usepackage{setspace}
\usepackage{scalerel}
\usepackage{subcaption}
\usepackage{tikz}
\usetikzlibrary{positioning}
\usepackage{verbatim}

\begin{document}

\preprint{APS/123-QED}

\title{Bayesian optimization to infer parameters in viscoelasticity}% Force line breaks with \\
%\thanks{A footnote to the article title}%

\author{Isaac Y. Miranda-Valdez}
 \email{isaac.mirandavaldez@aalto.fi}
\author{Tero M{\"a}kinen}%
\author{Juha Koivisto}%
\author{Mikko J. Alava}%

\affiliation{%
Department of Applied Physics, Aalto University, P.O. Box 15600, 00076 Aalto, Espoo, Finland}%

\date{\today}% It is always \today, today,
             %  but any date may be explicitly specified

\begin{abstract}
Inferring viscoelasticity parameters is a key challenge that often leads to non-unique solutions when fitting rheological data. In this context, we propose a machine learning approach that utilizes  Bayesian optimization for parameter inference during curve-fitting processes. To fit a viscoelastic model to rheological data, the Bayesian optimization maps the parameter values to a given error function. It then exploits the mapped space to identify parameter combinations that minimize the error. We compare the Bayesian optimization results to traditional fitting routines and demonstrate that our approach finds the fitting parameters in a less or similar number of iterations. Furthermore, it also creates a "white-box" and supervised framework for parameter estimation in linear viscoelasticity modeling.
%\begin{description}
%\item[Usage]
%Secondary publications and information retrieval purposes.
%\item[Structure]
%You may use the \texttt{description} environment to structure your abstract;
%use the optional argument of the \verb+\item+ command to give the category of each item. 
%\end{description}
\end{abstract}

%\keywords{Suggested keywords}%Use showkeys class option if keyword
                              %display desired
\maketitle

%\tableofcontents
%\linenumbers\relax % Commence numbering lines

Bayesian optimization (BO) is a probabilistic machine learning framework with broad applications in materials discovery and design~\citep{arróyave_khatamsaz_vela_couperthwaite_molkeri_singh_johnson_qian_srivastava_allaire_2022, torsti_mäkinen_bonfanti_koivisto_alava_2024, miranda-valdez_mäkinen_coffeng_päivänsalo_jannuzzi_viitanen_koivisto_alava_2025, makinen2024bayesian}. BO finds the extrema (maximum or minimum values) of costly objective functions using prior knowledge about a problem. Unlike traditional optimization techniques, which often rely on gradient information or exhaustive searches, BO actively guides sampling processes toward exploring and exploiting search spaces~\citep{frazier2018tutorialbayesianoptimization}. This approach is practical when a closed-form expression for the objective function is inaccessible, but one can gather (noisy) observations via simulations or experimentation. In this context, BO is particularly appealing as a framework to address the non-convex problem of inferring parameters when fitting viscoelastic models to rheological data.

Here, we focus on fractional order viscoelastic models, which compactly describe the power-law time-dependent behavior of soft materials~\citep{bonfanti_kaplan_charras_kabla_2020, song_holten-andersen_mckinley_2023}. These models have few parameters but involve complex numerical calculations, including e.g.~the Mittag--Leffler function~\citep{song_holten-andersen_mckinley_2023}. Fitting fractional models to rheological data is physically meaningful as it provides valuable material information, such as length and time scales~\citep{bantawa_keshavarz_geri_bouzid_divoux_mckinley_delgado_2023}. However, current fitting techniques often rely on manual parameter tuning, which may not effectively navigate local minima in the objective function landscape, leading to inaccurate parameter inference. Effectively addressing these local minima to find a global solution is crucial for the reliability of the fitted model.

In this article, we expand upon the Python package, \hyperlink{https://github.com/mirandi1/pyRheo}{\texttt{pyRheo}}, which we have recently developed for rheological modeling~\citep{mirandavaldez2024pyrheoopensourcepythonpackage}. 
Our focus is on utilizing BO to fit viscoelastic models to rheological data. The process consists of constructing a surrogate space with Gaussian Process Regression (GPR) and iteratively refining it with an acquisition function. 

We begin by importing a viscoelastic model from \texttt{pyRheo}, and the task is to identify the parameter values for this model that best fit the experimental dataset. To assess the convergence of the fitting process, we compute an error function. Using GPR, we map the error to its corresponding parameter values. An acquisition function then proposes new parameter values that are likely to minimize this error function. The search continues using the knowledge acquired from each iteration until the BO converges to the global minimum. The approach differs from normal Bayesian inference~\citep{freund_ewoldt_2015} for parameter estimation as we are not putting priors on the parameters and inferring the posterior distribution of the parameters, but instead use a Gaussian Process model for the residual errors and find the minimum of this error.\\

%\section{Results and discussion}

\begin{figure}[h]
    \centering
    \begin{tikzpicture}
        % First subfigure (Top left)
        \node (img1) at (0,1) {\includegraphics[width=0.47\linewidth]{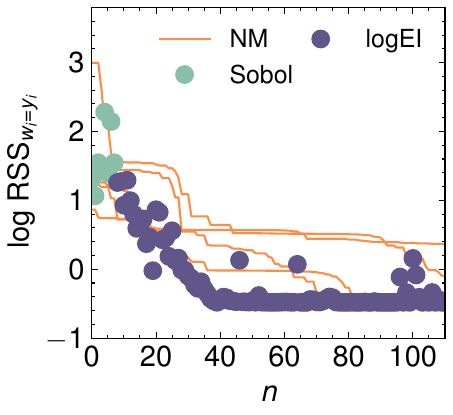}};
        % Add label 'a' to the upper left of img1
        \node[anchor=north west] at ([xshift=0.0cm, yshift=0.2cm]img1.north west) {\textbf{\Large a}};
        % Second subfigure (Bottom left)
        \node[anchor=north west] (img2) at ([yshift=0.0cm]img1.north east)  {\includegraphics[width=0.49\linewidth]{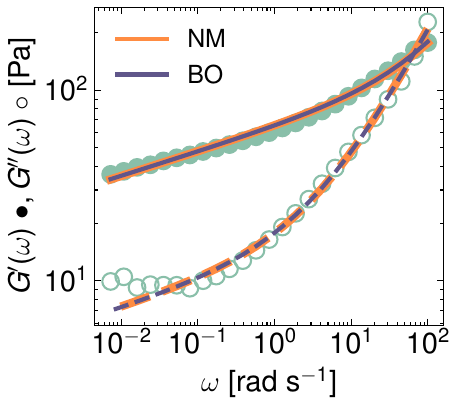}};
        \node[anchor=north west] (img3) at ([xshift=2.85cm, yshift=2.15cm]img2.south west)  {\includegraphics[width=0.160\linewidth]{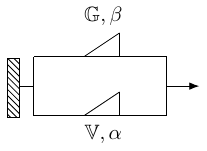}};
        % Add label 'b' to the upper left of img2
        \node[anchor=north west] at ([xshift=0.0cm, yshift=0.2cm]img2.north west) {\textbf{\Large b}};
        % Third subfigure (Top right)
        \node[anchor=north west] (img4) at ([yshift=0.0cm]img1.south west) {\includegraphics[width=0.98\linewidth]{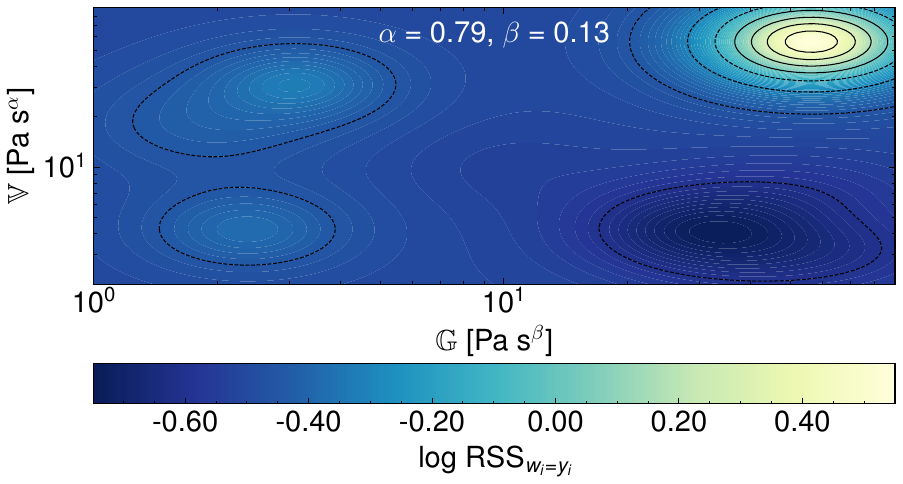}};
        % Add label 'c' to the upper left of img4
        \node[anchor=north west] at ([xshift=0.0cm, yshift=-0.14cm]img4.north west) {\textbf{\Large c}};
    \end{tikzpicture}
    \caption{Bayesian optimization (BO) to fit Fractional Kelvin-Voigt model (FKVM) to oscillation data measured for a chia gel. 
    {\bf a} Exploration-exploitation sequence showing the Sobol mapping of the error function, (weighted residual sum of squares $\text{RSS}_{w_{i}=y_i}$), %weighted with the function values $w_i=y_i$), 
    and its further exploitation using logarithmic Expected Improvement (logEI). The solid lines represent the minimization path followed by the Nelder-Mead (NM) algorithm initialized five times with random initial guesses. 
    {\bf b} Fit of the FKVM to the experimental model using the parameter values returned by the BO and the best result from the NM (solid line and dashed line correspond to storage and loss modulus, respectively). 
    {\bf c} Contour plot showing the final surrogate model for different values of $\mathbb{G}$ and $\mathbb{V}$ for fixed (optimal) $\alpha$ and~$\beta$.  }
    \label{fig:bo_oscillation_plots}
\end{figure}
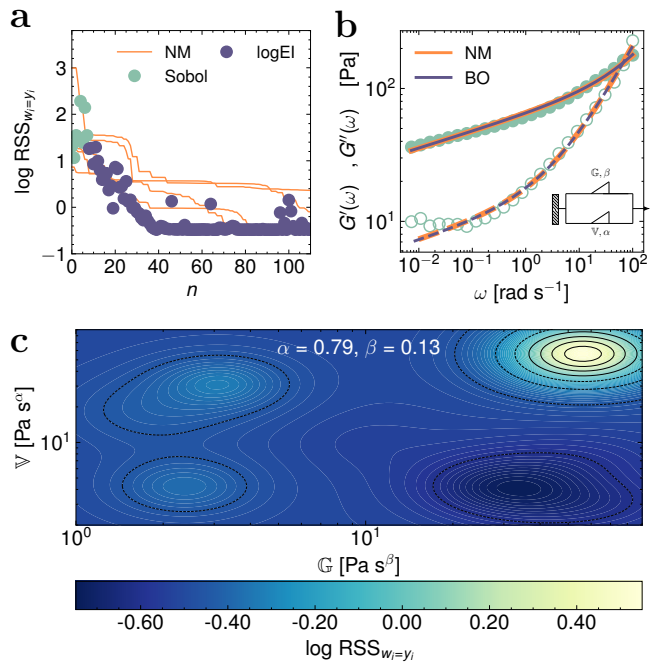

\begin{figure*}[!]
    \centering
    \begin{tikzpicture}
        % First subfigure (Top left)
        \node (img1) at (0,1) {\includegraphics[width=0.24\linewidth]{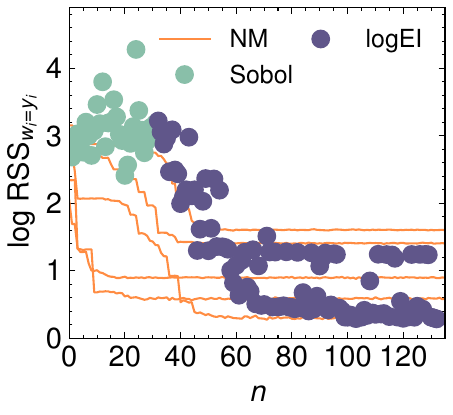}};
        % Add label 'a' to the upper left of img1
        \node[anchor=north west] at ([xshift=0.0cm, yshift=0.2cm]img1.north west) {\textbf{\Large a}};
        % Second subfigure (Bottom left)
        \node[anchor=north west] (img2) at ([yshift=0.0cm]img1.north east)  {\includegraphics[width=0.24\linewidth]{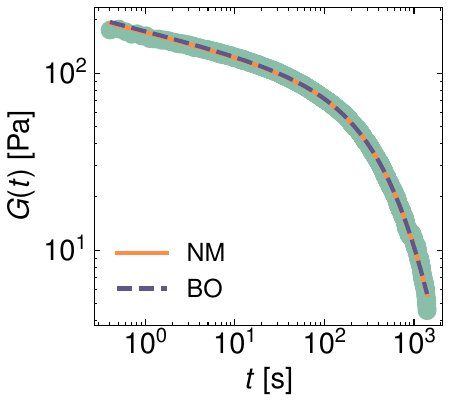}};
        \node[anchor=north west] (img3) at ([xshift=1.2cm, yshift=3.cm]img2.south west)  {\includegraphics[width=0.1\linewidth]{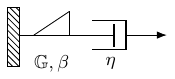}};
        % Add label 'b' to the upper left of img2
        \node[anchor=north west] at ([xshift=0.0cm, yshift=0.2cm]img2.north west) {\textbf{\Large b}};
        % Third subfigure (Top right)
        \node[anchor=north west] (img4) at ([yshift=0.2cm]img2.north east) {\includegraphics[width=0.49\linewidth]{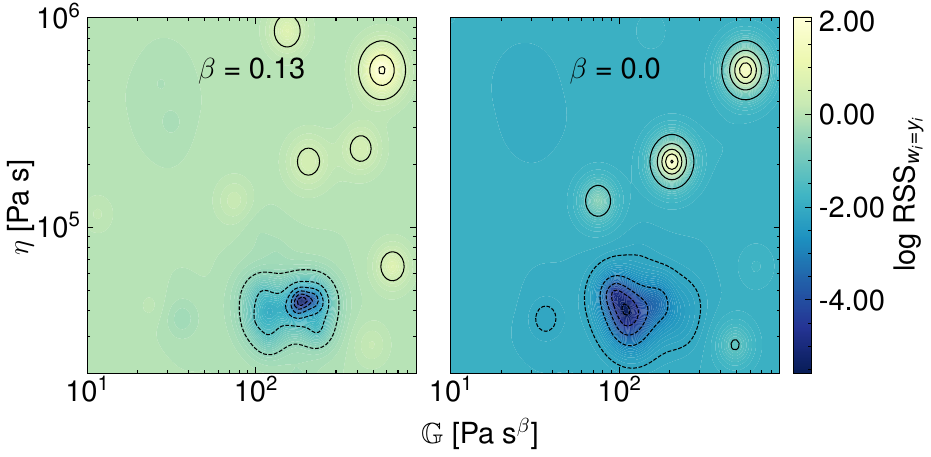}};
        % Add label 'c' to the upper left of img4
        \node[anchor=north west] at ([xshift=0.0cm, yshift=-0.0cm]img4.north west) {\textbf{\Large c}};
    \end{tikzpicture}
    \caption{Bayesian optimization (BO) to fit Fractional Maxwell Liquid model (FML) to the stress relaxation data measured for a shaving foam (experimental data reproduced from \citet{lavergne_sollich_trappe_2022}). 
    {\bf a}~Exploration-exploitation sequence showing the Sobol mapping of the error function (weighted residual sum of squares $\text{RSS}_{w_{i}=y_i}$), and its further exploitation using logarithmic Expected Improvement (logEI). The solid lines represent the minimization path followed by the Nelder-Mead (NM) algorithm initialized five times with random initial guesses. 
    {\bf b}~Fitting of the FML to the experimental data using the parameter values returned by the BO and the best result from the NM. 
    {\bf c}~Contour plots showing the final surrogate model for different values of $\mathbb{G}$ and $\eta$ for fixed $\beta$ values. We compare the BO solution with $\beta=0.13$ to the classical response of a Maxwell model when $\beta=0.0$.}
    \label{fig:bo_relaxation_plots}
\end{figure*}

Next, we present the results of using BO to fit data from oscillatory, relaxation, and creep tests. For oscillation experiments, we analyze the storage modulus ($G^{\prime}(\omega)$) and loss modulus ($G^{\prime \prime}(\omega)$) of a chia gel, whose viscoelastic behavior follows a Fractional Kelvin-Voigt model (FKVM)~\citep{mirandavaldez2024pyrheoopensourcepythonpackage}. Thus, we train a surrogate model that maps the FKVM parameter space $\mathcal{P}$ to an objective error space $\mathcal{E}$ using GPR, represented as a mapping $k: \mathcal{P} \to \mathcal{E}$, where $k$ is a Gaussian Process ($\mathcal{GP}$) with radial basis function (RBF) covariance kernel as implemented in \texttt{BoTorch}~\citep{balandat2020botorchframeworkefficientmontecarlo}. The surrogate model $\epsilon = k(p)$ (with $\epsilon \in \mathcal{E}$ and $p \in \mathcal{P}$) is trained by computing the constitutive equation of the FKVM $f(x_i; \theta)$ with a given combination of parameter values~$\theta$ and then recording the error $\epsilon$, weighted residual sum of squares (RSS$_{w_{i}}$), 
\begin{equation}
    \text{RSS}_{w_{i}} = \sum_{i=1}^{n} \left(\frac{y_i - f(x_i; \theta)}{w_i}\right)^2,
    \label{eq:supp_rsswi}
\end{equation}
between this computation and the experimental data. We give as weight $w_i = y_i$ to the RSS$_{w_{i}}$, where $y_i$ corresponds to each observation in the experimental dataset and $f$ to the viscoelastic model prediction. We note that it is feasible to use any choice of weight or even a different error function, as we demonstrate in the creep example. The Supplementary Information file details the BO workflow and related algorithms.

As Fig.~\ref{fig:bo_oscillation_plots}a shows, we start with an exploration phase of eight iterations ($n=8$) using a Sobol sequence~\cite{sobol1967distribution, owen1998scrambling} to initialize and train the surrogate model. The Sobol sequence samples the parameter space $\mathcal{P}$ evenly and improves the convergence rate of the acquisition function~\citep{balandat2020botorchframeworkefficientmontecarlo}. Furthermore, using a Sobol sequence during the exploration phase is computationally less demanding than employing an acquisition function. The second part in Fig.~\ref{fig:bo_oscillation_plots}a corresponds to the exploitation phase. We use logarithmic Expected Improvement (logEI) as recently proposed by \citet{ament2025unexpectedimprovementsexpectedimprovement}. logEI belongs to a new family of acquisition functions which substantially improve on the optimization performance of their canonical counterparts (i.e., EI). In the Supplementary Information section, we present an instance demonstrating the better performance of logEI than traditional EI. With logEI, we search for the combination of parameters that minimizes $\epsilon$,~\citep{sourroubille_miranda-valdez_mäkinen_koivisto_alava_2025, miranda-valdez_mäkinen_coffeng_päivänsalo_jannuzzi_viitanen_koivisto_alava_2025, ament2025unexpectedimprovementsexpectedimprovement}. We restrict the exploitation phase to a maximum of $n_{\rm{max}}=200$ with automatic stop criteria if no improvement seen after $n=100$. The logEI function actively guides the BO workflow by looking for the potential gains over the given design points $\mathbf{p} \in \mathcal{P}$ and exploits the search space with more gains. 

After roughly 50 guided acquisitions, the BO finds the "best" combination of parameters, 
\begin{equation}
    \mathbf{p} = \left[ \begin{matrix}
    \mathbb{G} = 65.2 \\ \mathbb{V} = 5.10  \\ \alpha = 0.79 \\ \beta =  0.13
\end{matrix} \right] \in \mathcal{P}.
\end{equation}
The blue solid and dashed lines depict the resulting fit in Fig.~\ref{fig:bo_oscillation_plots}b. Fig.~\ref{fig:bo_oscillation_plots}c shows the non-convex landscape built by the BO to find the optimal combination of parameters that minimizes RSS$_{w_{i}=y_i}$. There are a few considerations to implement in the BO workflow to build and exploit the landscape in Fig.~\ref{fig:bo_oscillation_plots}c effectively. The first is to consider the length scale ($l$) of the covariance kernel for the $\mathcal{GP}$, which determines the smoothness of the surrogate model. This is because large values of $l$ allow for smoother transitions between data points, whereas small $l$ values tend to assume that the function can change significantly over short distances.

Finding the optimal value for $l$ is crucial to prevent under and overfitting. To find an optimal $l$ for a $\mathcal{GP}$, we have shown for rheological data that the search space has to be built using logarithmic transformations in order to train the $\mathcal{GP}$, which account for the power-law rheology of soft materials~\citep{sourroubille_miranda-valdez_mäkinen_koivisto_alava_2025}. This approach has also been used by \citet{lennon_mckinley_swan_2023}. Accordingly, we build the surrogate model in Fig.~\ref{fig:bo_oscillation_plots}c by taking the $\log_{10}$ of $\mathbb{G}$ and $\mathbb{V}$. On the other hand, the fractional orders, $\alpha$ and $\beta$ do not need scaling. Similarly, since the error function can vary several orders of magnitudes from one observation to another, scaling is important to avoid underfitting the surrogate model. Therefore, to build the surrogate model in Fig.~\ref{fig:bo_oscillation_plots}c, we log-transformed $\mathcal{E}$. In addition, we subsequently scaled $\mathcal{E}$ to have zero mean and unit variance to improve the surrogate model convergence.

We translate the same BO workflow described for fitting oscillation data to the stress relaxation of shaving foam (experimental data reproduced from \citet{lavergne_sollich_trappe_2022}). We fit a Fractional Maxwell Liquid model (FML) to the relaxation modulus ($G(t)$) of the shaving foam. The stress relaxation function of the FML is expensive to compute because it includes the Mittag--Leffler function---an infinite sum of gamma functions~\citep{song_holten-andersen_mckinley_2023}. Therefore, one would like to avoid an exhaustive number of iterations to find the best fitting parameters. In Fig.~\ref{fig:bo_relaxation_plots}a, the BO workflow efficiently navigates the parameter space of the FML in 50 guided acquisitions and finds the best combination:
\begin{equation}
    \mathbf{p} = \left[ \begin{matrix}
    \mathbb{G} = 1.89\times10^{2}\\ \eta = 4.44\times10^{4}  \\ \beta =  0.13
\end{matrix} \right] \in \mathcal{P}.
\end{equation}
The resulting fitting using these parameters is depicted by the solid line in Fig.~\ref{fig:bo_relaxation_plots}b.

We build the landscape of the surrogate model in Fig.~\ref{fig:bo_oscillation_plots}c using noise-free observations. However, we can also train the surrogate model with noisy observations, which would typically account for the experimental error, which is the case of the surrogate model in Fig.~\ref{fig:bo_relaxation_plots}c and the reason why we do not see a "perfect" fit for the data at short times in Fig.~\ref{fig:bo_relaxation_plots}b. Without a known experimental error, we arbitrarily add a 1\% error to the FML function obtained from every iteration $n$. Including the effect of experimental error is a common need in rheological modeling~\citep{singh_soulages_ewoldt_2019, singh2022simultaneousfittingnonlinearlinear}. In the Supplementary Information accompanying this manuscript, we have included more examples of fittings done with BO for different materials and types of experiments. The scripts to compute the results of this manuscript are available as Jupyter Notebooks, which, together with the data, can be accessed from the GitHub~\url{https://github.com/mirandi1/pyRheo.git}.

% Comparison of results

We compare our BO approach to the conventional method used in rheology. For this comparison, we utilize the \texttt{minimize} function from SciPy~\citep{2020SciPy-NMeth} with the Nelder-Mead algorithm (NM)~\citep{nelder1968asimplexmethod}. The NM is a reliable method for low-dimensional and constrained minimization problems. Similar to BO, we use the NM to find the combination of parameters that minimizes the error function, RSS$_{w_{i}=y_i}$.

In our analysis of the datasets related to oscillation and relaxation, as shown in Fig.~\ref{fig:bo_oscillation_plots}a and Fig.~\ref{fig:bo_relaxation_plots}a, we observe that both the BO and NM require a similar number of iterations to reach the global minima. However, the NM is sensitive to the initial conditions. For example, in Fig.~\ref{fig:bo_oscillation_plots}a the NM can take up to 250 iterations to converge to the local minima depending on the initial guesses given whereas BO needs only 50. In both datasets, some attempts result in convergence to local minima when the NM is initialized with five random initial guesses. This issue is particularly pronounced in the stress relaxation example, which has a complex parameter landscape (depicted in Fig.~\ref{fig:bo_relaxation_plots}c) and thus only one of the five NM initializations found the global minima. 

\begin{figure*}[!]
    \centering
    \begin{tikzpicture}
        % First subfigure (Top left)
        \node (img1) at (0,1) {\includegraphics[width=0.24\linewidth]{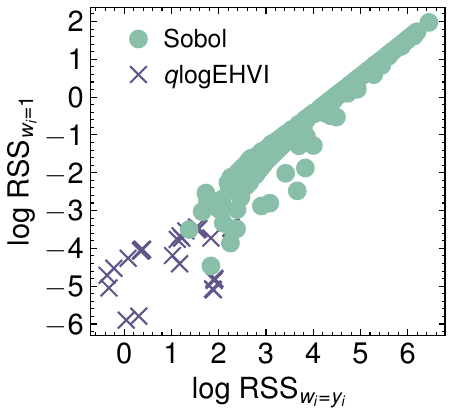}};
        % Add label 'a' to the upper left of img1
        \node[anchor=north west] at ([xshift=0.0cm, yshift=0.2cm]img1.north west) {\textbf{\Large a}};
        % Second subfigure (Bottom left)
        \node[anchor=north west] (img2) at ([yshift=0.0cm]img1.north east)  {\includegraphics[width=0.25\linewidth]{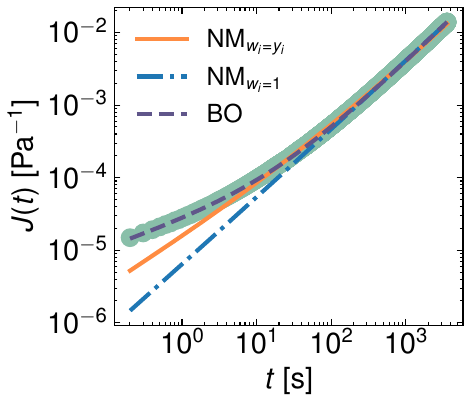}};
        \node[anchor=north west] (img3) at ([xshift=2.7cm, yshift=2cm]img2.south west)  {\includegraphics[width=0.1\linewidth]{fml_scheme.pdf}};
        % Add label 'b' to the upper left of img2
        \node[anchor=north west] at ([xshift=0.0cm, yshift=0.2cm]img2.north west) {\textbf{\Large b}};
        % Third subfigure (Top right)
        \node[anchor=north west] (img4) at ([yshift=0.2cm]img2.north east) {\includegraphics[width=0.24\linewidth]{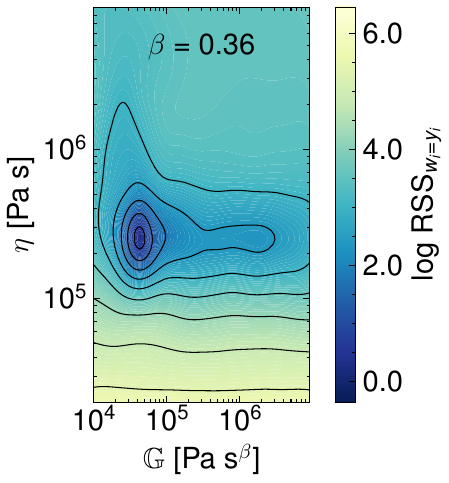}};
        % Add label 'c' to the upper left of img4
        \node[anchor=north west] at ([xshift=0.0cm, yshift=-0.0cm]img4.north west) {\textbf{\Large c}};
        % Third subfigure (Top right)
        \node[anchor=north west] (img5) at ([yshift=0.0cm]img4.north east) {\includegraphics[width=0.24\linewidth]{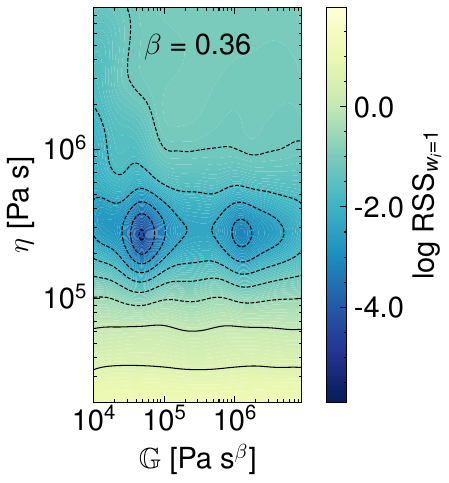}};
        % Add label 'c' to the upper left of img4
        \node[anchor=north west] at ([xshift=0.0cm, yshift=-0.0cm]img5.north west) {\textbf{\Large d}};
    \end{tikzpicture}
    \caption{Multi-objective Bayesian optimization (BO) to fit Fractional Maxwell Liquid model (FML) to creep compliance data measured for polystyrene melt at 175$^{\circ}$C (experimental data reproduced from \citet{ricarte_shanbhag_2024}). 
    {\bf a}~Exploration-exploitation sequence showing the Sobol mapping of the error function (weighted residual sum of squares RSS$_{w_{i}}$ with different weights $w_i = y_i$ and $w_i = 1$), and their further exploitation using $q$-Expected Hypervolume Improvement ($q$logEHVI). 
    {\bf b}~Fitting of the FML to the experimental model using the parameter values returned by the multi-objective BO and the best results from the NM.
    {\bf c}~Contour plot showing the final surrogate model for RSS$_{w_{i}=y_i}$ for different values of $\mathbb{G}$ and $\eta$ for the optimal $\beta$~value.
    {\bf d}~Same as panel c but for RSS$_{w_{i}=1}$.
    }
    \label{fig:bo_creep_plots}
\end{figure*}

In terms of computational efficiency, the BO approach requires slightly more resources to fit rheological data. For instance, the solution given for the oscillation dataset takes around an order of magnitude longer than the five NM minimizations (5.72~s \textit{vs.} 0.12~s on a desktop computer with the characteristics reported in the Supplementary Information). Nevertheless, the BO method excels when the viscoelastic model function is expensive to evaluate as in the case of the relaxation dataset. BO finds the fitting parameters plotted for Fig.~\ref{fig:bo_relaxation_plots}b in a similar time as the NM minimizations (7.76~s \textit{vs.} 2.51~s).

% Conclusion
Although we have focused on fitting fractional order viscoelastic models in this manuscript, the BO workflow can be adapted to solve other constrained and non-convex problems in soft matter and rheology. For example, this approach could be extended beyond rheology to other areas of soft matter research, such as polymer science, biomaterials, and colloidal systems, where modeling non-linear and complex behaviors is crucial. In particular, with the recent developments in machine learning algorithms, the approach presented here can be extended to multi-objective optimization. For instance, we can target to minimize the same error function in Eq.~\ref{eq:supp_rsswi} but with different weights that trade-off the areas with high and low absolute residuals. 

We present in Fig.~\ref{fig:bo_creep_plots} an example of minimizing the RSS$_{w_{i}}$ with different weights ($w_i =
y_i$ and $w_i = 1$) to fit the FML to the creep compliance ($J(t)$) of a polystyrene melt at 175$^{\circ}$C (experimental data reproduced from \citet{ricarte_shanbhag_2024}). As seen in Fig.~\ref{fig:bo_creep_plots}a, the multi-objective BO follows an approach similar to the one described in the oscillation and stress relaxation examples. We train two surrogate models that map the FML parameter space $\mathcal{P}$ to two objective error spaces $\mathcal{E}_1$ and $\mathcal{E}_2$ using GPR, represented as  $k: \mathcal{P} \to \mathcal{E}_1$ and $g: \mathcal{P} \to \mathcal{E}_2$, where $k$ and $g$ are $\mathcal{GP}$s with RBF covariance kernels. 

The process starts with a Sobol exploration phase with $n=256$. We refine the mapped space with an acquisition function ($n=50$), $q$-logExpected Hypervolume Improvement (qlogEHVI)~\citep{balandat2020botorchframeworkefficientmontecarlo}. It quantifies how much the combined hypervolume in $\mathcal{E}$ would improve if a new solution were sampled over the current Pareto front, given by the current best observations from the exploration phase. The solution obtained from the acquisition function is:
\begin{equation}
    \mathbf{p} = \left[ \begin{matrix}
    \mathbb{G} = 4.62\times10^{4}\\ \eta = 2.65\times10^{5}  \\ \beta =  0.36
\end{matrix} \right] \in \mathcal{P}.
\end{equation}
Fig.~\ref{fig:bo_creep_plots}b shows the fitting obtained using the multi-objective BO.

In the multi-objective BO, $n$ is high in the exploration phase. We observe that the multi-objective BO is computationally intensive, as it creates two surrogate models, as illustrated in Fig.~\ref{fig:bo_creep_plots}c and d, and thus is is cheaper to explore first the surrogate spaces and than exploit them with the acquisition function. The multi-objective BO approach may require more resources than one might be willing to allocate for fitting rheological data. However, as shown in Fig.~\ref{fig:bo_creep_plots}b, multi-objective BO achieves a lower RSS$_{w_{i}}$ (where $w_i = y_i$ and $w_i = 1$) compared to the NM method. Fig.~\ref{fig:bo_creep_plots}b presents the best results from five NM minimizations for each objective using different random initial guesses. The traditional minimization tends to struggle to describe the part of the creep compliance at short times where the function values are low. Therefore, optimizing both RSS$_{w_i}$ objectives with BO helps better describe the short-time data by balancing the small residuals given by the RSS$_{w_i}$ with $w_i = 1$ with the information of the RSS$_{w_i}$ weighted by function values ($w_i = y_i$), which can yield larger absolute residuals in the same time span.

In conclusion, we highlight BO as a machine learning approach to advance rheological modeling. BO offers a supervised approach to understanding the non-convex landscape that results from mapping the parameter space of a given viscoelastic model to error functions. The results exemplify how a strategic sampling method such as BO can lead to more informed fittings, which can account for experimental uncertainty. 
%We must note that the approach here presented is not related to Bayesian inference. Bayesian inference has been used in rheology to justify model selection based on the number of parameters and the \textit{a priori} uncertainty~\citep{freund_ewoldt_2015}. 
On the other hand, BO looks for the global minima using prior knowledge about the constrained optimization problem. We conclude that the values inferred with BO can be taken as the final combination of model parameters. A future step in this research is to implement the Bayesian Information Criterion~\cite{leonard2001bayesian, song_holten-andersen_mckinley_2023} in the objective function to minimize with BO, so the BO solution can be used to choose among different viscoelastic models.

\section*{Author Contributions}
{\bf Isaac Y. Miranda-Valdez:} Conceptualization, Methodology, Software, Formal analysis, Visualization, Investigation, Data curation, Writing -- Original draft, Writing -- Review \& Editing, Project administration, and Funding acquisition.
{\bf Tero Mäkinen:} Validation and Writing -- Review \& Editing.
{\bf Juha Koivisto:} Supervision, Validation, Writing -- Review \& Editing, Funding acquisition.
{\bf Mikko J. Alava:} Supervision, Validation, Writing -- Review \& Editing, Funding acquisition, Project administration.

\section*{Conflicts of interest}
There are no conflicts to declare.

\section*{Data availability}
The code, data analysis scripts, and data for this article are available at pyRheo's Github at \url{https://github.com/mirandi1/pyRheo/tree/main/demos/bayesian_optimization}.

\section*{Acknowledgements} 
I.M.V. acknowledges the Vilho, Yrjö, and Kalle Väisälä Foundation of the Finnish Academy of Science and Letters for personal funding. 
M.J.A. acknowledges funding from the Finnish Cultural Foundation.
M.J.A., J.K., T.M. and I.M.V. acknowledge funding from Business Finland (211909, 211989).
M.J.A. and J.K. acknowledge funding from FinnCERES flagship (151830423), Business Finland (211835), and Future Makers programs.
Aalto Science-IT project is acknowledged for computational resources.

\bibliography{main}% Produces the bibliography via BibTeX.

\end{document}

%%%%%%%%%
\begin{center}
    \fontfamily{phv}\selectfont\Large
    \textbf{Supplementary Information}
\end{center}
\normalsize

\hspace{2cm}

\noindent {\fontfamily{phv}\selectfont\Large \bf Bayesian optimization to infer parameters in viscoelasticity} \\
   
\noindent Isaac Y. Miranda-Valdez,\textit{$^{a,}$}$^{\ast}$ Tero M\"{a}kinen,\textit{$^{a}$}, Juha Koivisto,\textit{$^{a}$} and Mikko J. Alava,\textit{$^{a}$}\\

\noindent $^a$Complex Systems and Materials, Department of Applied Physics, Aalto University, P.O. Box 11000, FI-00076 Aalto, Espoo, Finland \\
\noindent $^\ast$ Corresponding author: isaac.mirandavaldez@aalto.fi\\

\noindent {\bf Abstract}  \\
This Supplementary Information accompanies the article \textit{Bayesian optimization to infer parameters in viscoelasticity}. In this file, we detail the single-objective and multi-objective Bayesian optimization (BO) workflows used in the main manuscript. Furthermore, we present more test examples where we have used BO to fit rheological data of different types and materials.\\

\begin{figure}[!h]
    \centering
    \includegraphics[width=1\linewidth]{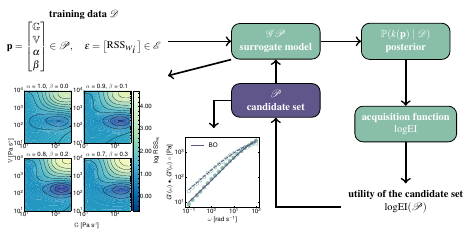}
    \label{fig:enter-label}
    \caption*{\textbf{Graphical abstract}: Bayesian optimization workflow using a Gaussian Process $\mathcal{GP}$ to map the parameters of a viscoelastic model to the error function, weighted residual sum of squares (RSS$_{w_i}$). Logarithmic Expected Improvement (logEI) exploits the $\mathcal{GP}$ to search for the candidate parameter combinations with more gains. The process is iterated multiple times until finding the "best" combination of parameters. The contour plot represents the final landscape mapped by the $\mathcal{GP}$, and the mechanical spectrum shows the experimental data to fit and the resulting fit using the Bayesian optimization candidate solution.}
\end{figure}

\clearpage
%%%%%%%%%%%%%%%%%%%%%%%%%%%%%%%%%%%%%%%%%%%%%%%%%%%%%%%%%%%%%%%%%%%%%%%%%%%%%

%\tableofcontents

%\clearpage

\section{Bayesian optimization}
\label{sec:workflow_methodology}

We use Bayesian optimization (BO) with an RBF covariance kernel, a popular choice for the Gaussian Process due to its flexibility and ability to model functions with varying degrees of smoothness. The RBF kernel is defined as:
\begin{equation}
    k(p, p') = \mu^2 \exp\left( - \frac{(p-p')^2}{2l^2} \right),
\end{equation}
where $\mu$ is the strength of the correlation or mean value, $l$ is the length scale, and $p-p'$ is the space feature between two input points.

We implement the BO workflows using the Python library \texttt{BoTorch}, which provides an interface for constructing BO problems~ \citep{balandat2020botorchframeworkefficientmontecarlo}. We use two BO frameworks, one based on single-objective optimization and the other on multi-objective optimization. For the viscoelastic models, we used the same notation as in \citet{song_holten-andersen_mckinley_2023}
and \citet{mirandavaldez2024pyrheoopensourcepythonpackage}. The reader can learn more about the viscoelastic models from the latter two publications.

\subsection{Single-objective Bayesian Optimization}

For the single-objective optimization, we aim to find the optimal parameters of a fractional viscoelastic model that fit the rheological data obtained from oscillation and stress relaxation tests. In the main manuscript, we use the data in oscillation measured for a chia gel as a case example, and we attempt to fit a Fractional Kelvin-Voigt model (FKVM). The equations of the FKVM for storage modulus $G^{\prime}(\omega)$ and loss modulus $G^{\prime\prime}(\omega)$ are given by
\begin{equation}
    \begin{aligned}
    G^{\prime}(\omega) &= \mathbb{V}\omega^{\alpha}\cos{(\frac{\pi}{2}\alpha)} + \mathbb{G}\omega^{\beta}\cos{(\frac{\pi}{2}\beta)} \\
    G^{\prime \prime}(\omega) &= \mathbb{V}\omega^{\alpha}\sin{(\frac{\pi}{2}\alpha)} + \mathbb{G}\omega^{\beta}\sin{(\frac{\pi}{2}\beta)},\\
    \end{aligned}
    \label{eq:oscillation_fvk}
\end{equation}
where $\mathbb{G}$, $\mathbb{V}$, $\alpha$, and $\beta$ are the FKVM parameters, and $\omega$ is the angular frequency. We use \texttt{pyRheo} to call the FKVM as a model object on Python~\citep{mirandavaldez2024pyrheoopensourcepythonpackage}.

The optimization workflow consists of the following steps:
\begin{enumerate}
    \item Initial exploration using Sobol sequence with $n$ acquisitions. The Sobol sequence assign values to the parameters $\mathbb{G}$, $\mathbb{V}$, $\alpha$, and $\beta$. It replaces these values in Eq.~\ref{eq:oscillation_fvk}. From each acquisition, we record the weighted residual sum of squares RSS$_{w_{i}}$ between the FKVM function obtained by replacing the parameter values and the experimental data.
    \item Subsequent exploitation to refine the search by minimizing RSS$_{w_{i}}$ between the model predictions and experimental data. The acquisition function used in this process is logarithmic Expected Improvement (logEI)~\citep{ament2025unexpectedimprovementsexpectedimprovement}, which is computed $n$ times to search for the minima.
\end{enumerate}
The RSS$_{w_{i}}$ is given by 
\begin{equation}
    \text{RSS}_{w_{i}} = \sum_{i=1}^{n} \left(\frac{y_i - f(x_i; \theta)}{w_i}\right)^2,
    \label{eq:supp_rsswi}
\end{equation}
where \(w_i = y_i\) , and \(f(x_i)\) are the viscoelastic model predicted values using the parameter values $\theta$.

In the case of the chia gel, the BO searches for the best combination of parameters in a constrained space defined as,
\[
\mathcal{P} = 
\begin{bmatrix}
    [0, 2] \\  % \log_{10}(G)
    [0, 2] \\  % \log_{10}(V)
    [0.5, 1] \\ % \alpha
    [0, 0.5]   % \beta
\end{bmatrix},
\]
where:
\begin{itemize}
    \item \( \log_{10}(\mathbb{G}) \): \([0, 2]\), the range for parameter \( \mathbb{G} \) in logarithmic scale.
    \item \( \log_{10}(\mathbb{V}) \): \([0, 2]\), the range for parameter \( \mathbb{V} \) in logarithmic scale.
    \item \( \alpha \): \([0.5, 1]\), the range for parameter \( \alpha \) in its original scale.
    \item \( \beta \): \([0, 0.5]\), the range for parameter \( \beta \) in its original scale.
\end{itemize}

The search space is denoted as \( \mathcal{P} \), while the corresponding outputs \( \mathcal{E} \) are represented by the weighted residual sum of squares, \( \text{RSS}_{w_i} \). To train the machine learning model, we standarize the scale of $\mathcal{E}$ by taking its $\log_{10}$ and further scaling it to have zero mean and unit variance. This defines the relationship where \( \mathcal{E} \) is obtained for each evaluated point \( \textbf{p} \in \mathcal{P} \) and modeled using a Gaussian Process $\mathcal{GP}$ represented as $k: \mathcal{P} \to \mathcal{E}$.

In the case of the stress relaxation of the shaving foam, the BO searches for the best combination of parameters in a constrained space defined as,
\[
\mathcal{P} = 
\begin{bmatrix}
    [1, 3] \\  % \log_{10}(G)
    [4, 6] \\  % \log_{10}(eta)
    [0, 0.5]   % \beta
\end{bmatrix},
\]
where:
\begin{itemize}
    \item \( \log_{10}(\mathbb{G}) \): \([1, 3]\), the range for parameter \( \mathbb{G} \) in logarithmic scale.
    \item \( \log_{10}(\eta) \): \([4, 6]\), the range for parameter \( \eta \) in logarithmic scale.
    \item \( \beta \): \([0, 0.5]\), the range for parameter \( \beta \) in its original scale.
\end{itemize}
This defines the relationship where \( \mathcal{E} \) is obtained for each evaluated point \( \textbf{p} \in \mathcal{P} \) by computing the stress relaxation function of the Fractional Maxwell Liquid model (FML). We use \texttt{pyRheo} to call the FML as a model object on Python, where the Mittag--Leffler function is $E_{a,b}(z)$~\citep{mirandavaldez2024pyrheoopensourcepythonpackage}. The relaxation modulus function of the FML is given by
\begin{equation}
    \begin{aligned}
    G(t) &= G_c \left(\frac{t}{\tau_c}\right)^{-\beta}E_{a,b}(z)\\
    \tau_c &= \left(\frac{\eta}{\mathbb{G}} \right)^{\frac{1}{1-\beta}}\\
    G_c &= \eta \tau_c^{-1} \\
    a &= 1 - \beta \\
    b &= 1 - \beta \\
    z &= -\left(\frac{t}{\tau_c}\right)^{1 - \beta}.
    \end{aligned}
    \label{eq:relaxation_fml}
\end{equation}
Similar to the chia example, we train a Gaussian Process $\mathcal{GP}$ represented as $k: \mathcal{P} \to \mathcal{E}$. One must note that the number of logEI acquisitions will change depending on the size of the search space (e.g., big search spaces requires a larger number of acquisitions than small spaces). %For example, changing the bounds of the shaving foam to
%\[
%\mathcal{P} = 
%\begin{bmatrix}
%    [1, 6] \\  % \log_{10}(G)
%    [1, 6] \\  % \log_{10}(eta)
%    [0, 0.5]   % \beta
%\end{bmatrix},
%\]
%increase the number of logEI acquisitions from $\sim$50 to $\sim$150.

\subsection{Multi-objective Optimization}
For the multi-objective optimization, we aim to minimize two objectives. We create two $\mathcal{GP}$ represented as $k: \mathcal{P} \to \mathcal{E}_1$ and $g: \mathcal{P} \to \mathcal{E}_2$. $\mathcal{E}_1$ and $\mathcal{E}_2$ are the weighted residual sum of squares RSS$_{w_{i}}$ with $w_i = y_1$ and $w_i = 1$, respectively. In the case of the polystyrene melt, the BO searches for the best combination of parameters in a constrained space defined as,
\[
\mathcal{P} = 
\begin{bmatrix}
    [4, 7] \\  % \log_{10}(G)
    [4, 7] \\  % \log_{10}(V)
    [0, 1]   % \beta
\end{bmatrix},
\]
where:
\begin{itemize}
    \item \( \log_{10}(\mathbb{G}) \): \([4, 7]\), the range for parameter \( \mathbb{G} \) in logarithmic scale.
    \item \( \log_{10}(\eta) \): \([4, 7]\), the range for parameter \( \eta \) in logarithmic scale.
    \item \( \beta \): \([0, 1]\), the range for parameter \( \beta \) in its original scale.
\end{itemize}
This defines the relationship where \( \mathcal{E}_1 \) and \( \mathcal{E}_2 \) are obtained for each evaluated point \( \textbf{p} \in \mathcal{P} \) by computing the creep compliance function $J(t)$ of the Fractional Maxwell Liquid model (FML)~\citep{song_holten-andersen_mckinley_2023, mirandavaldez2024pyrheoopensourcepythonpackage},
\begin{equation}
    J(t) = \frac{t}{\eta} + \frac{1}{\mathbb{G}}\frac{t^{\beta}}{\Gamma(1+\beta)},
    \label{eq:creep_fml}
\end{equation}
and modeled using two Gaussian Processes $k: \mathcal{P} \to \mathcal{E}_1$ and $g: \mathcal{P} \to \mathcal{E}_2$.

We trade-off between \( \mathcal{E}_1 \) and \( \mathcal{E}_2 \) using $q$-logExpected Hypervolume Improvement (qlogEHVI) as acquisition function~\citep{balandat2020botorchframeworkefficientmontecarlo}. The acquisition function looks in $\mathcal{P}$ for the combination of parameters that minimizes both objectives.

\clearpage
%%%%%%%%%%%%%%%%%%%%%%%%%%%%%%%%%%%%%%%%%%%%%%%%%%%%%%%%%%%%%%%%%%%%%%%%%%%%%

\section{More case examples}

In this section, we showcase more examples where we have used BO workflows to fit rheological data. In Supplementary~Fig.~\ref{fig:si_bo_creep_plots_ps190}, we use single-objective BO to fit the creep compliance data measured for a polystyrene melt at 190$^{\circ}$C (experimental data reproduced from \citet{ricarte_shanbhag_2024}).

\begin{figure*}[htbp]
    \centering
    \begin{tikzpicture}
        % First subfigure (Top left)
        \node (img1) at (0,1) {\includegraphics[width=0.24\linewidth]{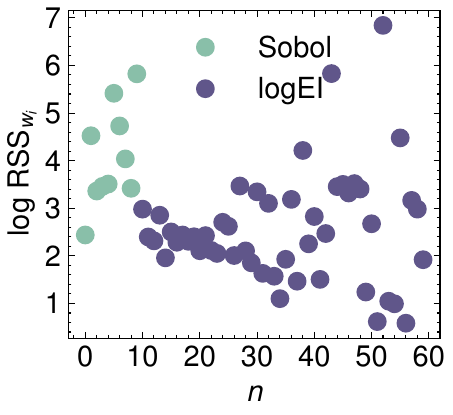}};
        % Add label 'a' to the upper left of img1
        \node[anchor=north west] at ([xshift=0.0cm, yshift=0.2cm]img1.north west) {\textbf{\Large a}};
        % Second subfigure (Bottom left)
        \node[anchor=north west] (img2) at ([yshift=0.2cm]img1.south west)  {\includegraphics[width=0.24\linewidth]{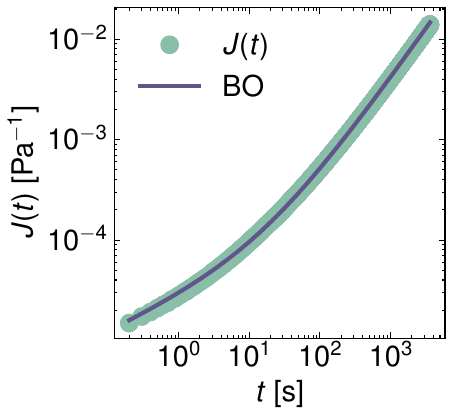}};
        \node[anchor=north west] (img3) at ([xshift=2.4cm, yshift=1.7cm]img2.south west)  {\includegraphics[width=0.09\linewidth]{fml_scheme.pdf}};
        % Add label 'b' to the upper left of img2
        \node[anchor=north west] at ([xshift=0.0cm, yshift=0.2cm]img2.north west) {\textbf{\Large b}};
        % Third subfigure (Top right)
        \node[anchor=north west] (img4) at ([yshift=0.35cm]img1.north east) {\includegraphics[width=0.74\linewidth]{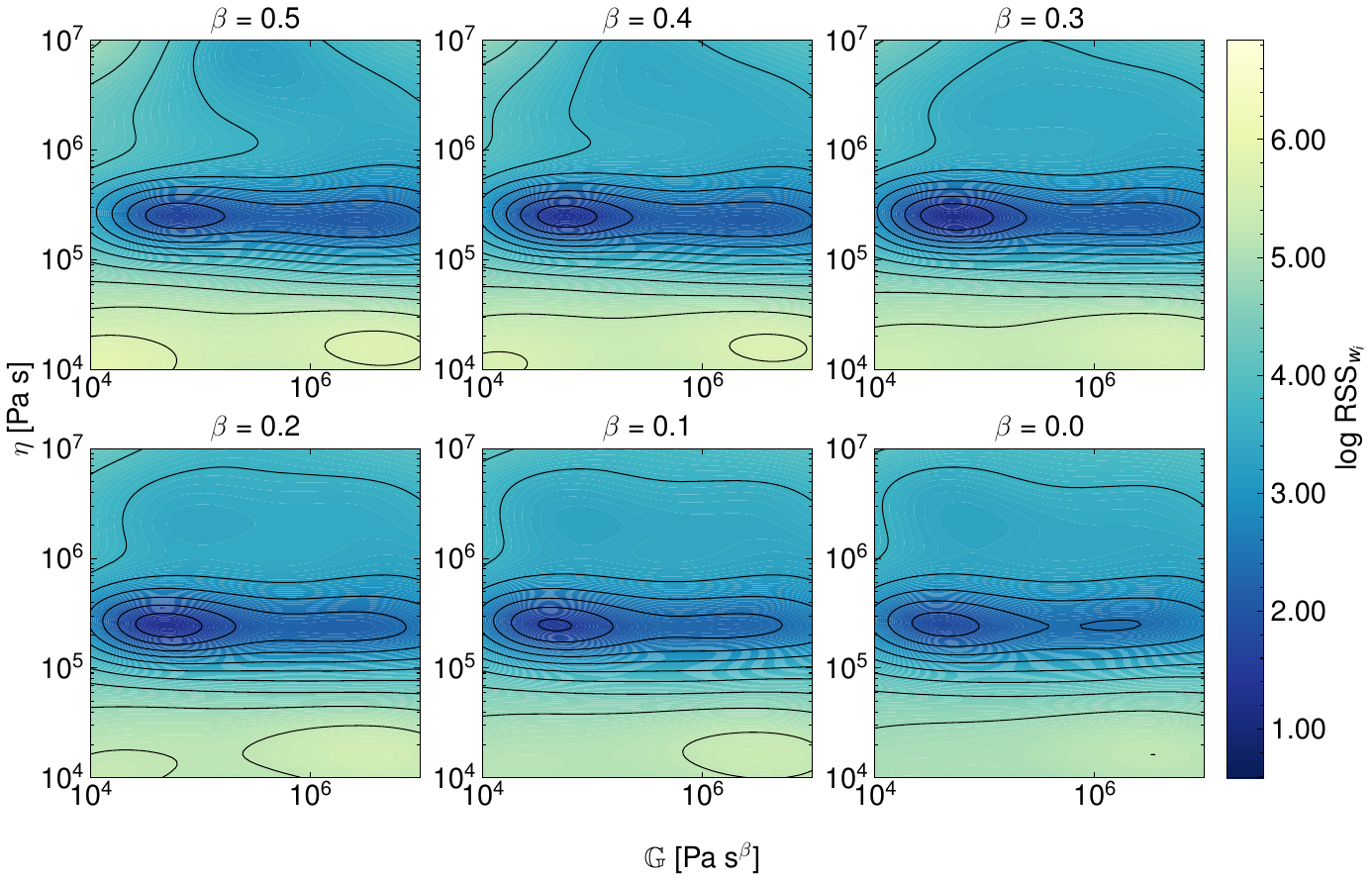}};
        % Add label 'c' to the upper left of img4
        \node[anchor=north west] at ([xshift=0.0cm, yshift=-0.14cm]img4.north west) {\textbf{\Large c}};
    \end{tikzpicture}
    \caption{Bayesian optimization (BO) to fit Fractional Maxwell Liquid model (FML) to the creep compliance data measured for a polystyrene melt at 190$^{\circ}$C (experimental data reproduced from \citet{ricarte_shanbhag_2024}). \textbf{a} Exploration-exploitation sequence showing the Sobol mapping of the error function, weighted residual sum of squares ($\text{RSS}_{w_{i}}$ weighted with the function values $w_i=y_i$)  in logarithm scale, and its further exploitation using logarithmic Expected Improvement (logEI). \textbf{b} Fitting of the FML to the experimental model using the parameter values returned by the BO. \textbf{c} Contour plots showing the final surrogate model for different values of $\mathbb{G}$ and $\eta$ for fixed $\beta$ values.}
    \label{fig:si_bo_creep_plots_ps190}
\end{figure*}

In Supplementary~Fig.~\ref{fig:si_bo_oscillation_plots_metal}, we use single-objective BO to fit the storage and loss modulus data measured for a metal-coordinating polymer network (experimental data obtained from \citet{lennon_mckinley_swan_2023} and reproduced from \citet{epstein_2019_metal}). 

\begin{figure*}[htbp]
    \centering
    \begin{tikzpicture}
        % First subfigure (Top left)
        \node (img1) at (0,1) {\includegraphics[width=0.24\linewidth]{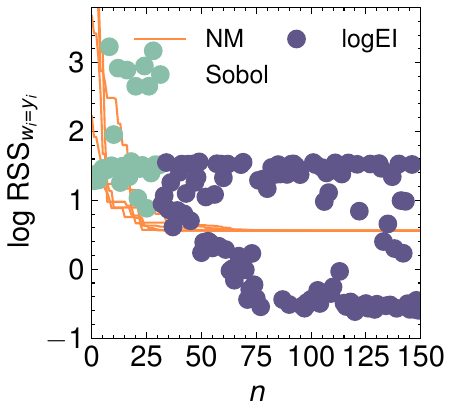}};
        % Add label 'a' to the upper left of img1
        \node[anchor=north west] at ([xshift=0.0cm, yshift=0.2cm]img1.north west) {\textbf{\Large a}};
        % Second subfigure (Bottom left)
        \node[anchor=north west] (img2) at ([yshift=0.2cm]img1.south west)  {\includegraphics[width=0.24\linewidth]{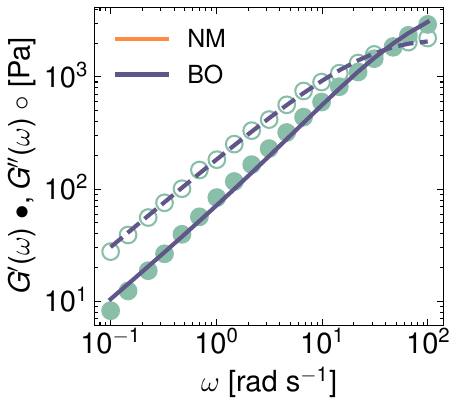}};
        \node[anchor=north west] (img3) at ([xshift=2.3cm, yshift=1.9cm]img2.south west)  {\includegraphics[width=0.09\linewidth]{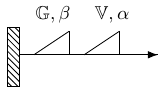}};
        % Add label 'b' to the upper left of img2
        \node[anchor=north west] at ([xshift=0.0cm, yshift=0.2cm]img2.north west) {\textbf{\Large b}};
        % Third subfigure (Top right)
        \node[anchor=north west] (img4) at ([yshift=0.35cm]img1.north east) {\includegraphics[width=0.74\linewidth]{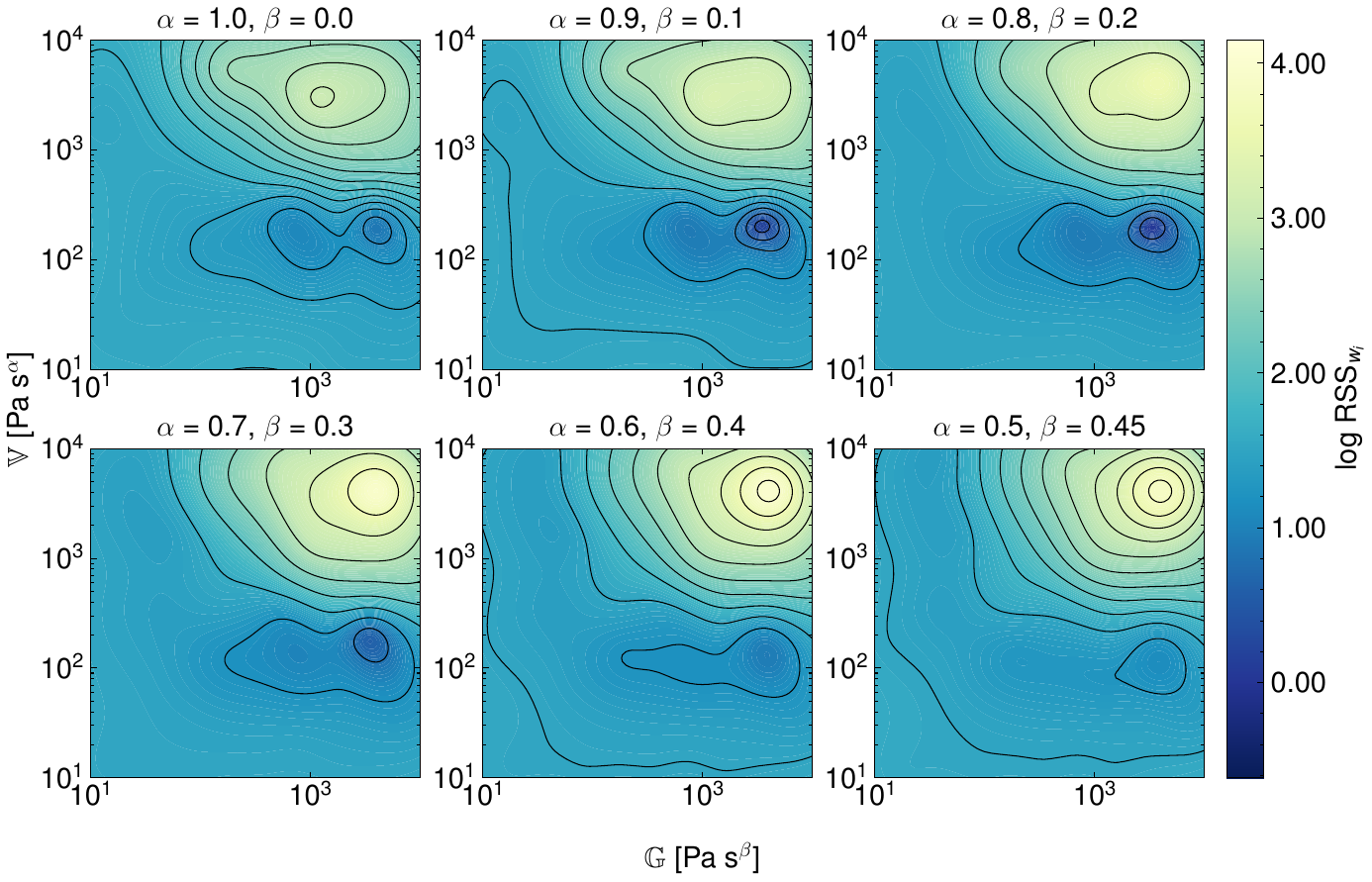}};
        % Add label 'c' to the upper left of img4
        \node[anchor=north west] at ([xshift=0.0cm, yshift=-0.14cm]img4.north west) {\textbf{\Large c}};
    \end{tikzpicture}
    \caption{Bayesian optimization (BO) to fit Fractional Maxwell Model (FMM) to the storage and loss modulus data measured for a metal-coordinating polymer networkd (experimental data obtained from \citet{lennon_mckinley_swan_2023} and reproduced from \citet{epstein_2019_metal}). \textbf{a} Exploration-exploitation sequence showing the Sobol mapping of the error function, weighted residual sum of squares ($\text{RSS}_{w_{i}}$ weighted with the function values $w_i=y_i$)  in logarithm scale, and its further exploitation using logarithmic Expected Improvement (logEI). \textbf{b} Fitting of the FMM to the experimental model using the parameter values returned by the BO. \textbf{c} Contour plots showing the final surrogate model for different values of $\mathbb{G}$ and $\mathbb{V}$ for fixed combinations of $\alpha$ and $\beta$ values.}
    \label{fig:si_bo_oscillation_plots_metal}
\end{figure*}

Finally, we compare the performance between the traditional Expected Improvement (EI) function and the logarithmic Expected Improvement (logEI) function. In Supplementary~Fig.~\ref{fig:si_bo_comparison_ei_logei}a,b, we compute the BO workflow using EI. In Supplementary~Fig.~\ref{fig:si_bo_comparison_ei_logei}c,d, we compute the same BO workflow but with logEI as the acquisition function. One can see that the EI function exhibits numerical instabilities when approaching the global minimum of the surrogate space. On the other hand, the logEI function performs a more effective exploitation of the surrogate space.

\begin{figure*}[htbp]
    \centering
    \begin{tikzpicture}
        % First subfigure (Top left)
        \node (img1) at (0,1) {\includegraphics[width=0.24\linewidth]{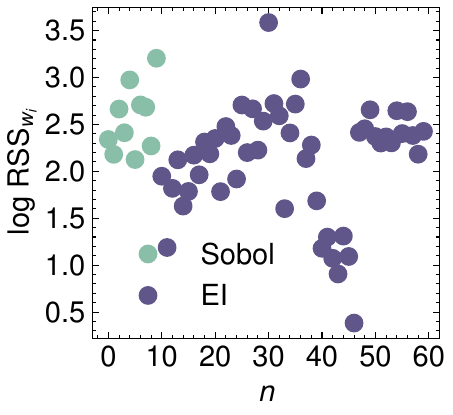}};
        % Add label 'a' to the upper left of img1
        \node[anchor=north west] at ([xshift=0.0cm, yshift=0.2cm]img1.north west) {\textbf{\Large a}};
        
        % Second subfigure (Bottom left)
        \node[anchor=north west] (img2) at ([yshift=0.2cm]img1.south west)  {\includegraphics[width=0.24\linewidth]{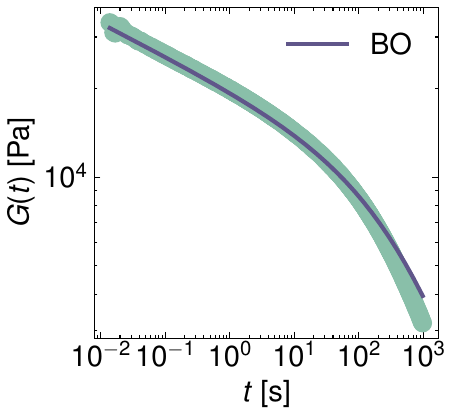}};
        \node[anchor=north west] (img3) at ([xshift=1cm, yshift=1.9cm]img2.south west)  {\includegraphics[width=0.09\linewidth]{fmm_scheme.pdf}};
        % Add label 'b' to the upper left of img2
        \node[anchor=north west] at ([xshift=0.0cm, yshift=0.2cm]img2.north west) {\textbf{\Large b}};

        % First subfigure (Top left)
        \node[anchor=north west] (img4) at (img1.north east)  {\includegraphics[width=0.24\linewidth]{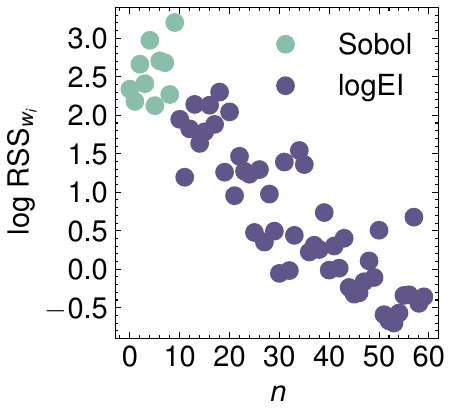}};
        % Add label 'b' to the upper left of img2
        \node[anchor=north west] at ([xshift=0.0cm, yshift=0.2cm]img4.north west) {\textbf{\Large c}};

        % Second subfigure (Bottom left)
        \node[anchor=north west] (img5) at ([yshift=0.0cm]img2.north east)  {\includegraphics[width=0.24\linewidth]{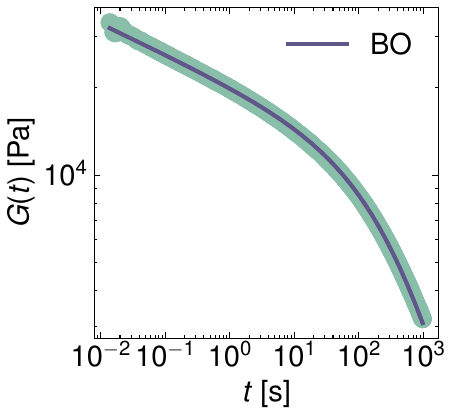}};
        \node[anchor=north west] (img6) at ([xshift=1cm, yshift=1.9cm]img5.south west)  {\includegraphics[width=0.09\linewidth]{fmm_scheme.pdf}};
        % Add label 'b' to the upper left of img2
        \node[anchor=north west] at ([xshift=0.0cm, yshift=0.2cm]img5.north west) {\textbf{\Large d}};
    \end{tikzpicture}
    \caption{Bayesian optimization (BO) to fit Fractional Maxwell Model (FML) to the relaxation modulus data of a fish muscle (experimental data reproduced from \citet{song_holten-andersen_mckinley_2023}). \textbf{a} Exploration-exploitation sequence showing the Sobol mapping of the error function, weighted residual sum of squares ($\text{RSS}_{w_{i}}$ weighted with the function values $w_i=y_i$)  in logarithm scale, and its further exploitation using logarithmic Expected Improvement (logEI). \textbf{b} Fitting of the FMM to the experimental model using the parameter values returned by the BO with EI. \textbf{c} The same as in \textbf{a} but exploitation with logarithmic Expected Improvement (logEI).  \textbf{d} Fitting of the FMM to the experimental model using the parameter values returned by the BO with logEI.}
    \label{fig:si_bo_comparison_ei_logei}
\end{figure*}

\clearpage
\bibliographystyle{references} 
\bibliography{supplementary}